\documentclass[appendixfloats]{emulateapj}
\usepackage{longtable}
\usepackage{graphicx}
\usepackage{graphics}
\usepackage{color}
\usepackage{epsfig}
\usepackage{rotating}
\usepackage{subfigure}
\usepackage{float}
\usepackage{url}
\usepackage{hyperref}
\usepackage{breakurl}
\usepackage{scalerel}
\usepackage[titletoc,title]{appendix}
\usepackage{relsize}

\shortauthors{J. T. Li et al.}
\shorttitle{Molecular Gas NGC~5908}

\begin{document}

\title{Molecular Gas of the Most Massive Spiral Galaxies I: a Case Study of NGC 5908}

\author{Jiang-Tao Li\altaffilmark{1}, Ping Zhou\altaffilmark{2}, Xuejian Jiang\altaffilmark{3}, Joel N. Bregman\altaffilmark{1}, Yu Gao\altaffilmark{3}} 

\altaffiltext{1}{Department of Astronomy, University of Michigan, 311 West Hall, 1085 S. University Ave, Ann Arbor, MI, 48109-1107, U.S.A.}

\altaffiltext{2}{Anton Pannekoek Institute for Astronomy, University of Amsterdam, Science Park 904, 1098 XH, Amsterdam, Netherlands}

\altaffiltext{3}{Purple Mountain Observatory/Key Lab of Radio Astronomy, Chinese Academy of Sciences, Nanjing 210034, China}


\nonumber

\begin{abstract}
We present IRAM 30m observations of molecular lines of CO and its isotopologues from the massive spiral galaxy NGC~5908 selected from the CGM-MASS sample. $^{12}$CO~$J=1-0$, $^{12}$CO~$J=2-1$, and $^{13}$CO~$J=1-0$ lines have been detected in most of the positions along the galactic disk. The total molecular gas mass of NGC~5908 is $\sim7\times10^9\rm~M_\odot$ and the total cool gas mass adding atomic hydrogen is $\sim1.3\times10^{10}\rm~M_\odot$, comparable to the upper limit of the mass of the X-ray emitting hot gas in the halo. Modeling the rotation curves constructed with all three CO lines indicates that NGC~5908 has a dark matter halo mass of $M_{\rm vir}\sim10^{13}\rm~M_{\rm \odot}$, putting it among the most massive isolated spiral galaxies. The $^{12}$CO/$^{13}$CO~$J=1-0$, $^{12}$CO~$J=2-1$/$J=1-0$ line ratios and the estimated molecular gas temperature all indicate normal but non-negligible star formation in this fairly gas-rich massive isolated spiral galaxy, consistent with the measured star formation intensity and surface densities. The galaxy is probably at an early evolutionary stage after a fast growth stage with mergers and/or starbursts, with plenty of leftover cool gas, relatively high SFR, low hot CGM cooling rate, and low X-ray emissivity.
\end{abstract}

\section{Introduction}\label{sec:Introduction}

The evolution of an isolated spiral galaxy is regulated by the cooling and accretion of the circum-galactic medium (CGM), the condensation of the cool gas in the disk and the follow-up star formation, as well as the feedback of energy and metal-enriched material by young stars and supernovae (SNe) back into the interstellar medium (ISM) and the CGM. In these processes, molecular gas in the galactic disk acts as a link for gas cycling between the CGM (via cooling and accretion) and the star formation (by providing fuel) (e.g., \citealt{Kennicutt12,Bolatto13} and references therein). Therefore, measuring the overall content and spatial distribution of molecular gas in an isolated spiral galaxy plays a key role in understanding how it builds up the stellar content of the galaxy. 

A spiral galaxy is not expected to survive a few major mergers, so the most massive galaxies in the local universe are always giant ellipticals in the center of clusters of galaxies. However, there exist extremely massive spiral galaxies with stellar mass a few times of the Milky Way (MW) in the local universe. These galaxies often have specific star formation rates (${\rm sSFR}={\rm SFR}/M_*$, where SFR and $M_*$ are the star formation rate and stellar mass of the galaxy) comparable to or lower than that of the MW, so it is impossible to build up their huge stellar mass with just continuous star formation at the current rate (e.g., \citealt{Li17}). Galaxy archeology with globular clusters (GCs) reveals that there are at least two stellar populations in massive spirals, associated with the disk and bulge respectively, with the disk component typically younger and with higher metallicity (e.g., \citealt{Cezario13}). This bimodality in the age and metallicity of GCs indicates that the major part of the disk of these galaxies may form after the merger and starburst stage when the galaxy swallowed most of the companions and became dominant in the dark matter halo. This scenario is further supported by the discovery of superluminous ($M_*=0.3-3.4\times10^{11}\rm~M_\odot$) spiral galaxies with active star formation (${\rm SFR}=5-65\rm~M_\odot~yr^{-1}$), diverse environments and merger stages at $z\sim0.1-0.3$ \citep{Ogle16}. These galaxies may be the progenitors of the local massive, isolated, quiescent spiral galaxies which represent the cleanest cases to witness the gentle growth of galactic disks without major merger and extensive star formation.

The above scenario indicates that the massive spiral galaxies observed in the local universe may have the major gas supply cutoff and the star formation stopped (also see discussions in \citealt{Schawinski14}). In this case, the leftover cold gas is quite informative in our understanding of the star formation, gas budget, and feedback efficiency in the post-starburst stage. We need to answer a few key questions to understand the formation of these extremely massive spiral galaxies, such as: (1) What is the amount and spatial distribution of the molecular gas? Do they maintain a regular, enhanced, or reduced star formation in the post starburst stage? (2) Is the gas circulation between the galaxy and CGM sufficient to maintain the observed amount of cold gas and star formation? (3) How does the existence of cold gas affect the feedback efficiency?

The Circum-Galactic Medium of MASsive Spirals (CGM-MASS) is a project studying the multi-phase CGM around the most massive (rotation velocity $v_{\rm rot}>300\rm~km~s^{-1}$, stellar mass $M_*\gtrsim1.5\times10^{11}\rm~M_\odot$) isolated spiral galaxies in the local universe \citep{Li16b,Li17,Li18}. The sample includes five galaxies observed in an \emph{XMM-Newton} AO-13 large program plus a few galaxies with high quality archival X-ray data. All of the sample galaxies appear to be less active in star formation than the star formation main sequence of low redshift galaxies (e.g., \citealt{Chang15}). None of them are in dense environments such as galaxy clusters. In order to study the molecular gas properties of these massive isolated quiescent spiral galaxies, we obtained deep IRAM 30m observations of the molecular lines of seven of the CGM-MASS galaxies visible in the northern sky in the 2016B, 2017A, and 2018B semesters with a total observing time of 25.2~hours. 

In this paper, we focus on the initial results from the IRAM 30m observations of NGC~5908, which has the highest quality data and is the nearest galaxy in the CGM-MASS sample (\citealt{Li16b,Li17}; see basic parameters of NGC~5908 in Table~\ref{table:NGC5908}). NGC~5908 is a highly-inclined massive spiral galaxy. We adopt a SFR estimated from the \emph{WISE} $22\rm~\mu m$ data ($3.8\pm0.1\rm~M_\odot~yr^{-1}$; \citealt{Li17}), which is $\approx43\%$ of what is obtained from the low-resolution \emph{IRAS} data \citep{Li16b}, and is consistent with the measured radio flux \citep{Condon02} and the radio-IR relation of edge-on galaxies (e.g., \citealt{Li16a}). The resultant specific SFR and SFR surface density are ${\rm SFR}/M_*=0.15\rm~M_\odot~yr^{-1}/(10^{10}\rm~M_\odot)$ and $\Sigma_{\rm SFR}=3\times10^{-3}\rm~M_\odot~yr^{-1}~kpc^{-2}$ (assuming the radius of the star forming disk is $r=20\rm~kpc$), respectively, both comparable to that of the MW \citep{Robitaille10,McMillan11}. For comparison, typical starburst galaxies have far larger ${\rm SFR}/M_*$ and $\Sigma_{\rm SFR}$ of $\gtrsim1\rm~M_\odot~yr^{-1}/(10^{10}\rm~M_\odot)$ and $\gtrsim10^{-2}\rm~M_\odot~yr^{-1}~kpc^{-2}$ (e.g., \citealt{Li13a,Li13b,Wang16}). Therefore, we regard NGC~5908 as a star formation inactive galaxy.
In the present paper, we will present data reduction in \S\ref{sec:datareduction}, main results and discussions in \S\ref{sec:resultsdiscussion}, and summarize the major conclusions in \S\ref{sec:summary}. Errors are quoted at 1~$\sigma$ level throughout the paper.

\begin{table}
\vspace{-0.in}
\begin{center}
\caption{Parameters of NGC~5908.} 
\footnotesize
\begin{tabular}{lcccccccccccccc}
\hline\hline
Parameters & Value & Reference \\
\hline
Type & SA(s)b & 1 \\
Distance & 51.9 Mpc & 1 \\
redshift & 0.011028 & 1 \\
scale & $1^{\prime\prime}=0.252\rm~kpc$ & 1 \\
inclination & $65.31^\circ$ & 2 \\
$v_{\rm rot}$ & $347\rm~km~s^{-1}$ & 3 \\
$M_{\rm HI}$ & $5.9\times10^9\rm~M_\odot$ & 4 \\
$M_*$ & $2.56_{-0.15}^{+0.02}\times10^{11}\rm~M_\odot$ & 6 \\
SFR & $3.81\pm0.09\rm~M_\odot~yr^{-1}$ & 6 \\
$M_{\rm 200}$ & $8.1\times10^{12}\rm~M_\odot$ & 5 \\
$r_{\rm 200}$ & 417~kpc & 5 \\
$kT$ & $0.38_{-0.09}^{+0.64}\rm~keV$ & 5,6 \\
$L_{\rm X,r<0.1r_{200}}$ & $7.2_{-2.3}^{+2.9}\times10^{39}\rm~ergs~s^{-1}$ & 5,6 \\
$r_{\rm cool}$ & $13.9_{-6.4}^{+4.9}\rm~kpc$ & 5,6 \\
$\dot{M}_{\rm cool,r<r_{cool}}$ & $0.37(<1.55)\rm~M_\odot~yr^{-1}$ & 5,6 \\ 
$M_{\rm hot,r<r_{200}}$ & $1.4_{-0.6}^{+3.3}\times10^{10}f^{1/2}\rm~M_\odot$ & 5,6 \\ 
\hline
$M_{\rm H_2,^{12}CO}$ & $(8.3\pm0.4)\times10^9\rm~M_\odot$ & 7 \\
$M_{\rm H_2,^{13}CO}$ & $(5.2\pm0.4)\times10^9\rm~M_\odot$ & 7 \\
$M_{\rm H_2,average}$ & $(6.8\pm1.7)\times10^9\rm~M_\odot$ & 7 \\
$\mathcal{R}_{\rm \mathsmaller{\mathsmaller{\frac{^{12}CO_{\rm 10}}{^{13}CO_{\rm 10}}}},median}$ & $7.0\pm1.7$ & 7 \\
$\mathcal{R}_{\rm \mathsmaller{\mathsmaller{\frac{^{12}CO_{\rm 21}}{^{12}CO_{\rm 10}}}},median}$ & $0.43\pm0.11$ & 7 \\
$\tau_{\rm ^{13}CO,median}$ & $0.16\pm0.03$ & 7 \\
$T_{\rm K,median}$ & $32\pm9\rm~K$ & 7 \\
$M_{\rm vir}$ & $(1.4\pm0.1)\times10^{12}\rm~M_\odot$ & 7 \\
$\Sigma_{\rm H_2}$ & $5.4\pm1.4\rm~M_\odot~pc^{-2}$ & 7 \\
$\Sigma_{\rm H_2+HI}$ & $10.1\pm1.4\rm~M_\odot~pc^{-2}$ & 7 \\
\hline\hline
\end{tabular}\label{table:NGC5908}
\end{center}
$v_{\rm rot}$ and $M_{\rm HI}$ are the inclination corrected rotation velocity and the atomic hydrogren mass obtained from the \ion{H}{1} 21-cm line observations. $M_*$ is the stellar mass estimated from the K-band magnitude. SFR is the star formation rate estimated from the \emph{WISE} $22\rm~\mu m$ data. $M_{\rm 200}$ is the dark matter halo mass (with density 200 times of the critical density of the Universe) estimated from $v_{\rm rot}$ with the NFW model (\citealt{Navarro96}). $r_{\rm 200}$ is the radius of the dark matter halo. $kT$ is the hot gas temperature. $L_{\rm X,r<0.1r_{200}}$ is the 0.5-2~keV soft X-ray luminosity of the hot gas component measured within $0.1r_{\rm 200}$. $r_{\rm cool}$ is the cooling radius defined as where the radiative cooling timescale equals to 10~Gyr. $\dot{M}_{\rm cool,r<r_{cool}}$ is the integrated radiative cooling rate calculated within $r_{\rm cool}$. $M_{\rm hot,r<r_{200}}$ is the total hot gas mass within $r_{\rm 200}$. All the hot gas parameters are estimated based on \emph{XMM-Newton} observations \citep{Li16b,Li17}. $M_{\rm H_2,^{12}CO}$ and $M_{\rm H_2,^{13}CO}$ are the molecular gas mass measured from the $^{12}$CO and $^{13}$CO~$J=1-0$ lines, respectively, while $M_{\rm H_2,average}$ is the average value of them. $\mathcal{R}_{\rm \mathsmaller{\mathsmaller{\frac{^{12}CO_{\rm 10}}{^{13}CO_{\rm 10}}}},median}$, $\mathcal{R}_{\rm \mathsmaller{\mathsmaller{\frac{^{12}CO_{\rm 21}}{^{12}CO_{\rm 10}}}},median}$, $\tau_{\rm ^{13}CO,median}$, $T_{\rm K,median}$ are the median values of the $^{12}$CO~$J=1-0$ to $^{13}$CO~$J=1-0$ ratio, the $^{12}$CO~$J=2-1$ to $^{12}$CO~$J=1-0$ ratio, the optical depth of $^{13}$CO~$J=1-0$, and the kinetic temperature of the gas measured at different positions along the disk. $M_{\rm vir}$ is the virial mass of the dark matter halo measured from the CO line rotation curves. $\Sigma_{\rm H_2}$ and $\Sigma_{\rm H_2+HI}$ are the surface density of the molecular or total cold gas, assuming the radius of the cold gas disk is $20\rm~kpc$.\\

References in this table:\\
1. NED (\url{https://ned.ipac.caltech.edu})\\
2. HyperLeda (\url{http://leda.univ-lyon1.fr})\\
3. \citet{vanMoorsel82}\\
4. \citet{Springob05}\\
5. \citet{Li16b}\\
6. \citet{Li17}\\
7. This paper
\end{table}

\section{Data Reduction}\label{sec:datareduction}

\begin{figure}
\begin{center}
\epsfig{figure=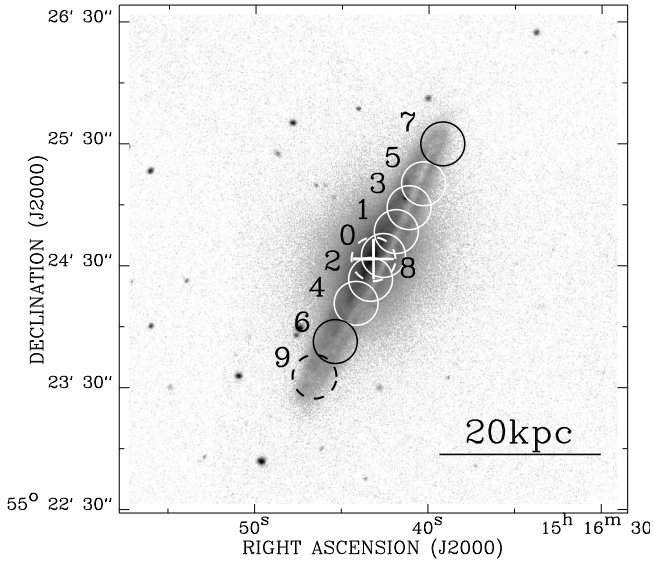,width=0.48\textwidth,angle=0, clip=}
\caption{SDSS r-band image of the central $4^\prime\times4^\prime$ region of NGC~5908. The circles are the location of the IRAM 30m beams with the size indicating the beam size at $^{12}$CO~$J=1-0$ band. The black solid circles have twice the exposure time as the white solid ones. The two dashed circles (``8'' and ``9'') are taken in the 2018B semester, with ``9'' has an exposure time comparable to the black solid circles while ``8'' has much shorter exposure time (see text for details). Most of the observations are along the dustlane, except for position ``8'' which is on the galactic center and slightly offset from the dustlane. The scale bar has a length of 20~kpc assuming a distance of $d=51.9\rm~Mpc$.
}\label{fig:IRAMbeam}
\end{center}
\end{figure}

The IRAM 30m observations of NGC~5908 was taken in 2016B (on July 26, 2016; Project ID 062-16; PI: Jiang-Tao Li) and 2018B semesters (on August 21, 2018; Project ID 063-18; PI: Jiang-Tao Li), respectively. The observations were taken with the Eight MIxer Receiver (EMIR) in Wobbler switching (WSW) mode in 2016B and in position switching (PSW) mode in 2018B, with the E90/E230 combination covering the CO $J=1-0$ and $J=2-1$ lines simultaneously. The fast Fourier transform spectrometer (FTS) backend was used, which provides a frequency resolution of $200\rm~kHz$. The half power beam width (HPBW) of IRAM 30m is $\sim10.7^{\prime\prime}$ at 230~GHz and $21.3^{\prime\prime}$ at 115~GHz. The 2016B observations were taken from eight positions along the dustlane of this highly inclined disk galaxy (Fig.~\ref{fig:IRAMbeam}). For the inner regions (white solid circles in Fig.~\ref{fig:IRAMbeam}), the exposure time is $30.6\rm~minutes$ for each position, while for the outermost regions at the edge of the dustlane (black solid circles in Fig.~\ref{fig:IRAMbeam}), we adopted $62\rm~minutes$ for each position to achieve a better sensitivity. We only observed two positions in the 2018B observations: the nuclear region (position ``8'' in Fig.~\ref{fig:IRAMbeam}) has an exposure time of $9.8\rm~minutes$, and the outer region at the edge of the dustlane (position ``9'' in Fig.~\ref{fig:IRAMbeam}) has an exposure time of $68.8\rm~minutes$. The 2018B observations did not cover the $^{13}$CO $J=1-0$ line at the redshift of NGC~5908.

\begin{figure*}
\begin{center}
\epsfig{figure=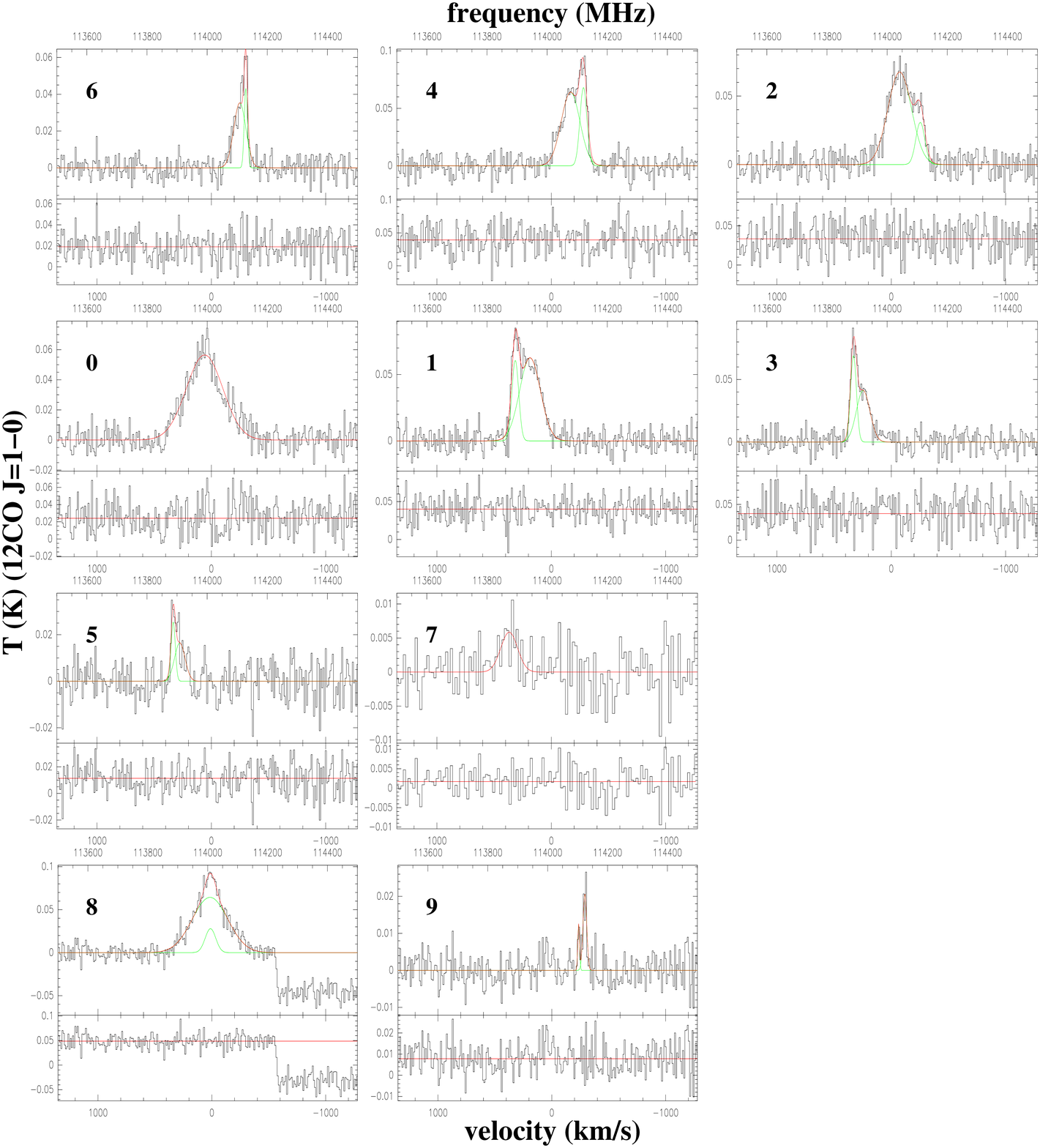,width=1.0\textwidth,angle=0, clip=}
\caption{$^{12}$CO~$J=1-0$ spectra of the beams shown in Fig.~\ref{fig:IRAMbeam} after subtracting a linear baseline. The poor baseline of ``8'' and ``9'' are caused by the poor weather conditions in the 2018B observations. The y-axis of the upper half of each panel is the main beam temperature after the main beam and forward efficiency corrections, while the lower half of each panel is the residual after subtracting the best-fit gaussian lines but adding the baseline. 
All the spectra have been binned to a velocity resolution of $\approx10\rm~km~s^{-1}$ (binned 20), expect for region 7, which has been binned to a resolution of $\approx20\rm~km~s^{-1}$ (binned 40). The spectra are fitted with a 1-gauss or 2-gauss model plus a 1-degree polynomial baseline. The red curve is the best-fit model, while the green curves, if present, are the two gaussian components. The zero velocity frequency is set to the systematic velocity of the galaxy ($3306\rm~km~s^{-1}$) and the horizontal range of different panels are the same ($113500-114500\rm~MHz$ for the upper axis, or $\sim-1351-+1278\rm~km~s^{-1}$ for the lower axis).}\label{fig:12CO10spec}
\end{center}
\end{figure*}

\begin{figure*}
\begin{center}
\epsfig{figure=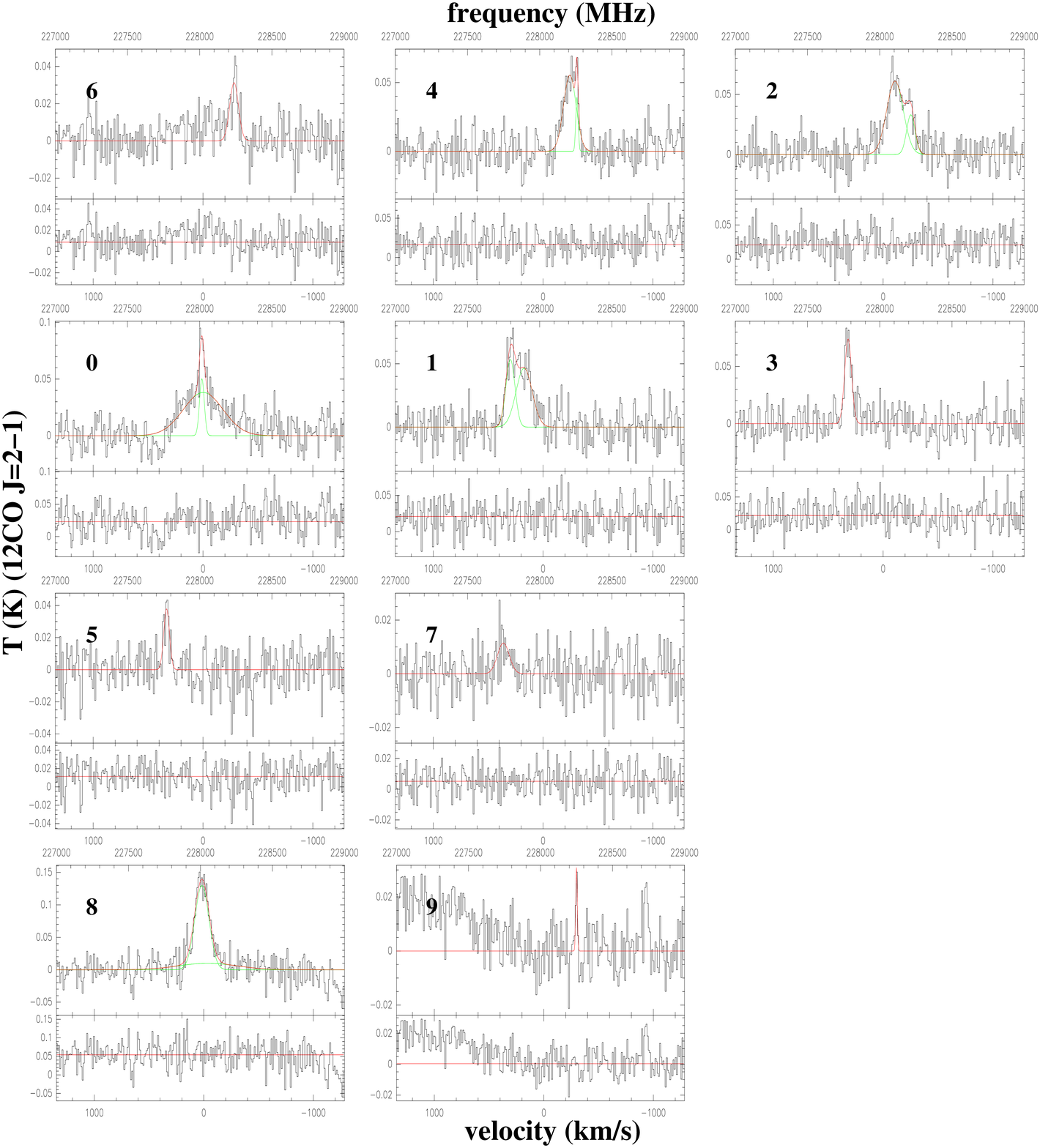,width=1.0\textwidth,angle=0, clip=}
\caption{Similar as Fig.~\ref{fig:12CO10spec}, but for the $^{12}$CO~$J=2-1$ spectra. All the spectra have been binned to a velocity resolution of $\approx10\rm~km~s^{-1}$ (binned 40). The horizontal range of different panels are the same ($227000-229000\rm~MHz$ for the upper axis, or $\sim-1345-+1284\rm~km~s^{-1}$ for the lower axis).}\label{fig:12CO21spec}
\end{center}
\end{figure*}

\begin{figure*}
\begin{center}
\epsfig{figure=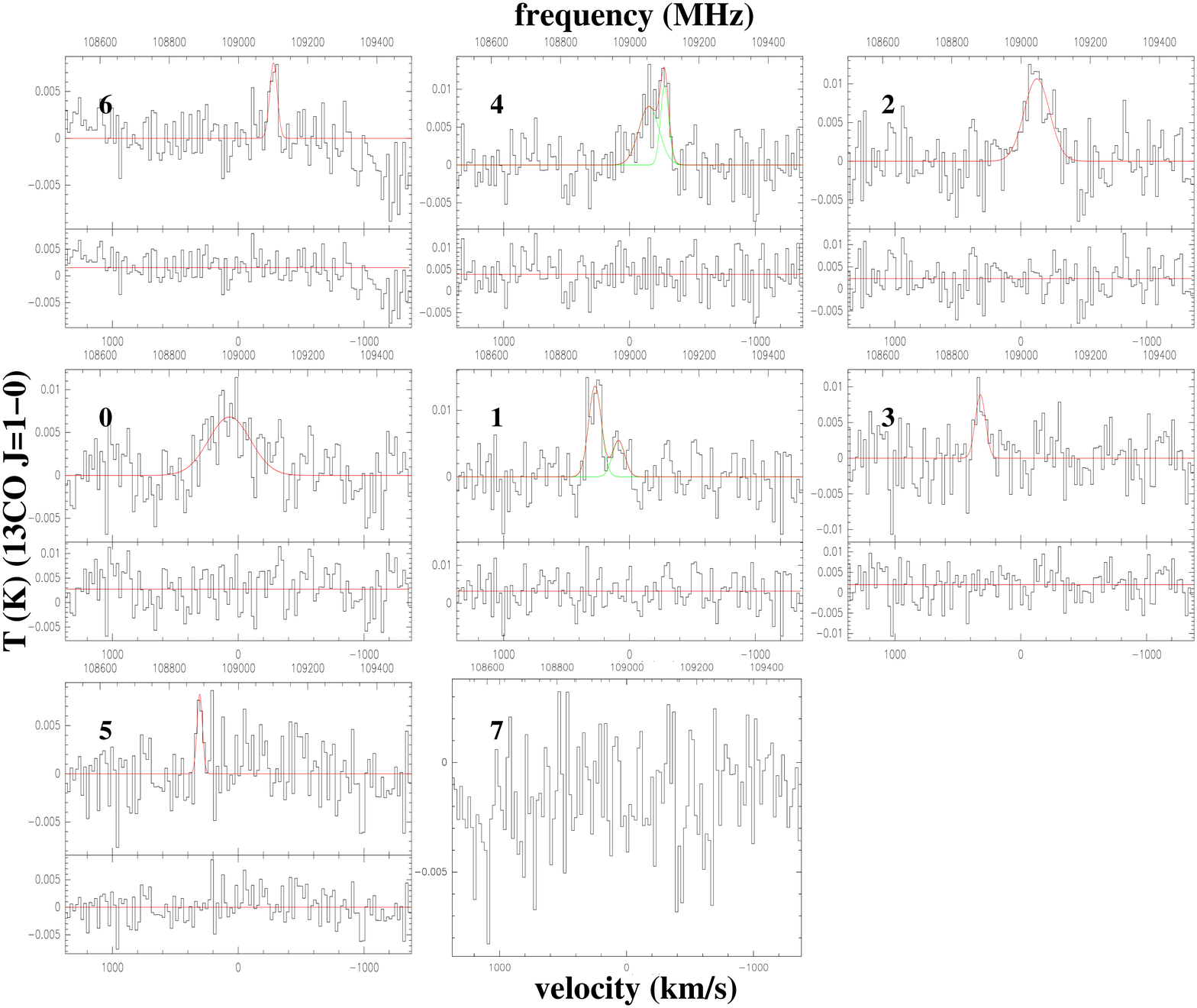,width=1.0\textwidth,angle=0, clip=}
\caption{Similar as Fig.~\ref{fig:12CO10spec}, but for the $^{13}$CO~$J=1-0$ spectra. All the spectra have been binned to a velocity resolution of $\approx20\rm~km~s^{-1}$ (binned 40). We do not detect any line at position 7. The horizontal range of different panels are the same ($108500-109500\rm~MHz$ for the upper axis, or $\sim-1373-+1377\rm~km~s^{-1}$ for the lower axis).}\label{fig:13CO10spec}
\end{center}
\end{figure*}

$^{12}$CO~$J=1-0$ (rest frame frequency $\nu_{\rm rest}=115.271\rm~GHz$; Fig.~\ref{fig:12CO10spec}), $^{12}$CO~$J=2-1$ ($\nu_{\rm rest}=230.538\rm~GHz$; Fig.~\ref{fig:12CO21spec}), and $^{13}$CO~$J=1-0$ lines ($\nu_{\rm rest}=110.201\rm~GHz$; Fig.~\ref{fig:13CO10spec}) have been detected at most of the positions (except for $^{13}$CO~$J=1-0$ at region~7 and the two positions 8 and 9 observed in 2018B) around the redshift of the galaxy. The velocity resolution was typically rebinned to $(10-20)\rm~km~s^{-1}$ to detect and well resolve the emission lines.

We calibrate and reduce the data following the standard steps with the GILDAS/CLASS package (\url{http://www.iram.fr/IRAMFR/GILDAS}; \citealt{Pety05}). The flux was calibrated during the observations using the standard sources. We correct the integrated line intensity ($I$, in unit of $\rm K~km~s^{-1}$, integrate the antenna temperature over a given velocity range) with the main beam efficiency using the recommended value of $B_{\rm eff}=78\%$ at 115~GHz for the $J=1-0$ transitions and 59\% at 230~GHz for the $J=2-1$ transition on the IRAM website (\url{http://www.iram.es/IRAMES/mainWiki/Iram30mEfficiencies}). We also adopt the recommended forward efficiencies of $F_{\rm eff}=94\%$ at 115~GHz for the $J=1-0$ transitions and 92\% at 230~GHz for the $J=2-1$ transition from the same website.

\begin{figure*}
\begin{center}
\epsfig{figure=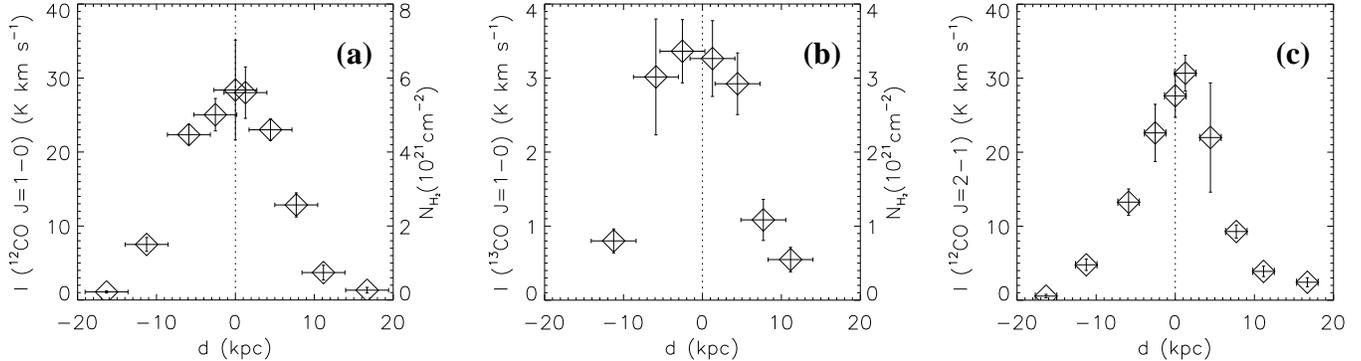,width=1.0\textwidth,angle=0, clip=}
\caption{The main beam and forward efficiency corrected integrated line intensity of $^{12}$CO~$J=1-0$ (a), $^{13}$CO~$J=1-0$ (b), and $^{12}$CO~$J=2-1$ (c) along the galactic disk. The x-axis is the distance along the galactic plane from the galactic center. The vertical line in each panel marks the location of the galactic center. We also convert the $^{12}$CO~$J=1-0$ and $^{13}$CO~$J=1-0$ intensity to the molecular gas column density $N_{\rm H_2}$ in (a) and (b).
}\label{fig:LineAreaProfile}
\end{center}
\end{figure*}

The emission lines are fitted with either a 1-gauss or 2-gauss model plus a linear baseline determined typically at $|v|<1400\rm~km~s^{-1}$ from the systematic velocity of the galaxy. The frequency windows used to fit the baseline are determined manually for each spectrum (windows are not plotted in Figs.~\ref{fig:12CO10spec}, \ref{fig:12CO21spec}, \ref{fig:13CO10spec}), after masking channels covering the emission lines or with significantly non-linear baselines (e.g., $v\lesssim-500\rm~km~s^{-1}$ in position-8 of Fig.~\ref{fig:12CO10spec}, $v\gtrsim400\rm~km~s^{-1}$ in position-9 of Fig.~\ref{fig:12CO21spec}, $|v|\gtrsim1000\rm~km~s^{-1}$ in position-6 of Fig.~\ref{fig:13CO10spec}, etc.). The number of gaussian lines used in the fit is determined by whether or not the data can be well fitted with a single gaussian model. In many regions, in addition to a broad component with a typical line width of $(100-400)\rm~km~s^{-1}$, there is often an additional narrow component with a line width of a few tens of $\rm~km~s^{-1}$. This decomposition may lead to the larger errors of some parameters measured in the inner region of the galaxy (e.g., Fig.~\ref{fig:LineAreaProfile}). In other regions, however, a single broad or narrow component can fit the data well. 

The broad and narrow components are often significantly shifted from each other. We therefore calculate an average line velocity by weighting the two components with their velocity integrated intensity. This averaged line velocity $v$ will be used to construct rotation curves in \S\ref{subsec:rotationcurve}. All these directly measured parameters are summarized in Table~\ref{table:COlinepara}. The corrected integrated line intensities are presented in Fig.~\ref{fig:LineAreaProfile}, while the rotation curves constructed with the averaged line velocities will be discussed in \S\ref{subsec:rotationcurve}. We do not list the width of the CO lines, which is not directly used in the paper, and in some cases largely affected by the decomposing of different components. In general, the largest width of the broad component around the center of the galaxy is $\sim400\rm~km~s^{-1}$ (Figs.~\ref{fig:12CO10spec}, \ref{fig:12CO21spec}, \ref{fig:13CO10spec}). If the line is dynamically broadened, such a large line width indicates a deep gravitational potential, as will be discussed in \S\ref{subsec:rotationcurve}.

\begin{figure*}
\begin{center}
\epsfig{figure=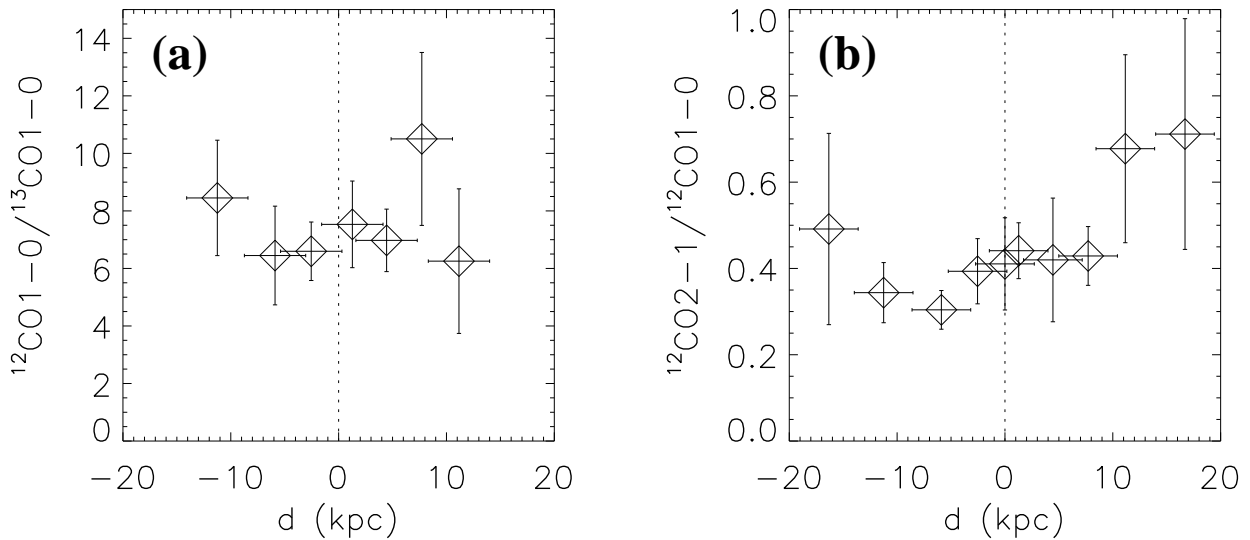,width=1.0\textwidth,angle=0, clip=}
\caption{(a) $^{12}$CO~$J=1-0$/$^{13}$CO~$J=1-0$ line ratio. (b) $^{12}$CO~$J=2-1$/$^{12}$CO~$J=1-0$ line ratio. Both line ratios have been corrected for beam dilution. The x-axis is the distance along the galactic plane from the galactic center.}\label{fig:RatioProfile}
\end{center}
\end{figure*}

For the ratio between different emission lines, we need to correct for the different beam dilution at different frequencies. NGC~5908 has been observed in CO~$J=1-0$ by CARMA as part of the EDGE-CALIFA Survey \citep{Bolatto17}. The CO~$J=1-0$ emission in the inner region of the galactic disk (radial extension comparable to the dustlane) is resolved and typically smaller than the IRAM 30m beam size in the vertical direction. The projected molecular gas distribution in NGC~5908 is therefore neither a point-like source nor a 2-D extended source. We assume the distribution of the molecular gas in NGC~5908 is ``disk'' like, i.e., continuous along the galactic disk while less extended than the main beam size of IRAM 30m in the vertical direction, consistent with the CARMA CO~$J=1-0$ image. We then construct 1-D ``images'' of different CO lines from the isolated IRAM 30m pointings. These CO images have a vertical extension equal to the size of the IRAM main beam at the corresponding frequencies. As we have chosen a separation of different pointings to be roughly equal to the IRAM main beam size at the CO~$J=2-1$ line, the intensity distributions along the galactic disk at all the frequencies of interest are continuous. We then convolve the higher frequency line images to the same resolution as the lower frequency ones when calculating their ratios. Compared to the directly measured line ratios, the measured beam dilution correction factor for the $^{12}$CO~$J=1-0$/$^{13}$CO~$J=1-0$ ratio is typically 0.87-0.95, while it is typically 0.38-0.65 for the $^{12}$CO~$J=2-1$/$^{12}$CO~$J=1-0$ ratio. The corrected line ratios are listed in Table~\ref{table:COlinepara} and plotted against the radial distance in Fig.~\ref{fig:RatioProfile}. 

\section{Results and Discussion}\label{sec:resultsdiscussion}

\subsection{Column density distribution and total mass of molecular gas}\label{subsec:IntensityNH2}

We first derive the molecular gas column density $N_{\rm H_2}$ from the main beam and forward efficiency corrected, integrated CO line intensity $I_{\rm CO}$ using the conversion factor defined as $X\equiv N_{\rm H_2}/I_{\rm CO}$, where $N_{\rm H_2}$ is in unit of $10^{21}\rm~cm^{-2}$ and $I_{\rm CO}$ is in unit of $\rm K~km~s^{-1}$. Both the ${\rm ^{12}CO}$ and ${\rm ^{13}CO}$~$J=1-0$ lines can be used to measure $N_{\rm H_2}$, but the conversion factor $X$ has large systematic uncertainties. We adopt an empirical $X_{{\rm ^{12}CO}~J=1-0}=0.2\times10^{21}\rm~cm^{-2}/(K~km~s^{-1})$ (e.g., \citealt{Nishiyama01,Shetty11,Bolatto13}), which however, may be highly variable in extreme physical conditions and is often underestimated with only low-$J$ CO transitions (e.g., \citealt{Papadopoulos12b}). The $^{12}$CO~$J=1-0$ line, owing to the high abundance of ${\rm ^{12}C}$, may be highly affected by the optical depth effect. On the other hand, the abundance of ${\rm ^{13}CO}$ and the physical state of molecular gas (in diffuse or dense clouds) both show significant variations across the disk of many nearby galaxies, resulting in a ${\rm ^{13}CO}$-to-$N_{\rm H_2}$ conversion factor $X_{{\rm ^{13}CO}~J=1-0}=1.0\times10^{21}\rm~cm^{-2}/(K~km~s^{-1})$ with a standard deviation of a factor of two (e.g., \citealt{Cormier18}].
In general, it is often suggested that $X_{\rm CO}$ decreases with increasing metallicity (e.g., \citealt{Schruba12}). A lower $X_{\rm CO}$ is also found in starburst galaxies or in the central bright regions with high stellar surface density in some galaxies (see a review in \citealt{Bolatto13}). However, none of these processes seem to highly affect the $X_{\rm CO}$ in a massive non-starburst galaxy such as NGC~5908. In this paper, we present $N_{\rm H_2}$ measured from both the ${\rm ^{12}CO}$ and ${\rm ^{13}CO}$~$J=1-0$ lines with the conversion factors of $X_{{\rm ^{12}CO}~J=1-0}=0.2\times10^{21}\rm~cm^{-2}/(K~km~s^{-1})$ and $X_{{\rm ^{13}CO}~J=1-0}=1.0\times10^{21}\rm~cm^{-2}/(K~km~s^{-1})$. The results are summarized in Table~\ref{table:COlinepara}.

Radial distribution of $N_{\rm H_2}$ as derived above is plotted together with the integrated intensities of the $^{12}$CO~$J=1-0$, $^{13}$CO~$J=1-0$, and $^{12}$CO~$J=2-1$ lines in Fig.~\ref{fig:LineAreaProfile}. We caution that the derived $N_{\rm H_2}$ depends on the filling factor $f$ of the CO emission region within the beam as $f^{-1}$. In highly inclined galaxies with most of the molecular gas distributed within a thin disk, the apparent vertical thickness of the molecular disk may be smaller than the IRAM 30m beam size (supported by the spatially resolved $^{12}$CO~$J=1-0$ image from \citealt{Bolatto17}), and $f$ may be significantly less than 1, so the derived $N_{\rm H_2}$ may just represent a lower limit of the real molecular gas column density.

Integrating the $N_{\rm H_2}$ profile and assuming the vertical extension of the molecular gas disk equals to the main beam size of IRAM 30m at the corresponding frequencies, we can obtain the total molecular gas mass of NGC~5908 to be $M_{\rm H_2}=8.26\pm0.35\times10^{9}\rm~M_\odot$ ($5.16\pm0.42\times10^{9}\rm~M_\odot$) based on ${\rm ^{12}CO}~J=1-0$ (${\rm ^{13}CO}~J=1-0$). Since $N_{\rm H_2}$ adopted in calculating $M_{\rm H_2}$ is an average value within the beam, this derived $M_{\rm H_2}$ does \emph{not} depend on the filling factor $f$.

\subsection{CO line ratios and physical conditions of molecular gas}\label{subsec:RatioPara}

\begin{figure*}
\begin{center}
\epsfig{figure=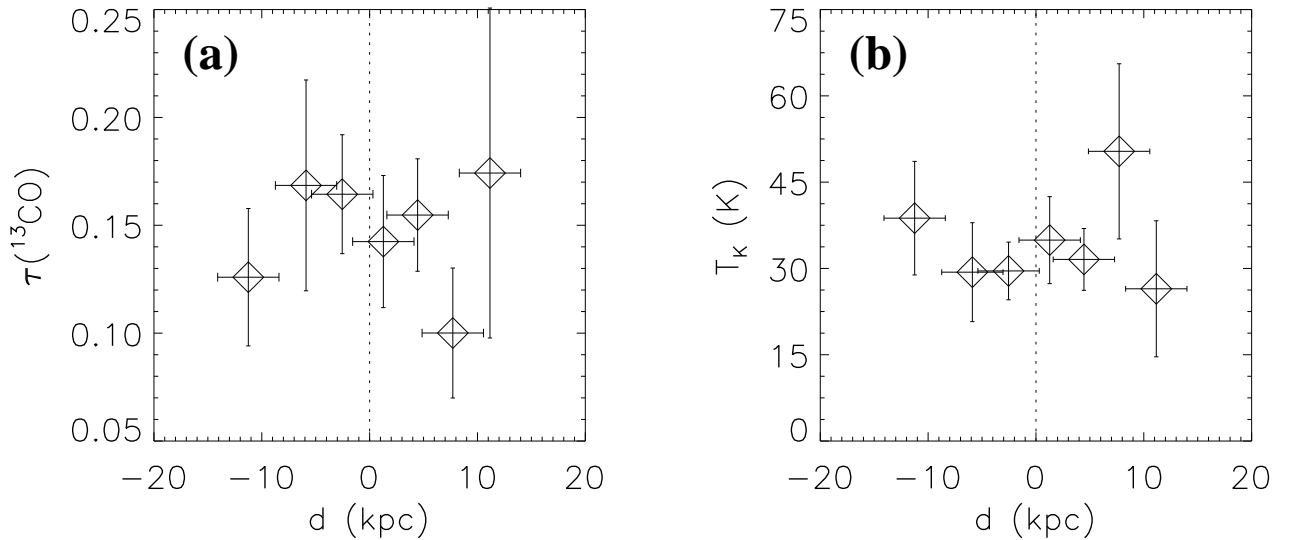,width=1.0\textwidth,angle=0, clip=}
\caption{(a) Optical depth of $^{13}$CO ($\tau_{\rm ^{13}CO}$). (b) Kinetic temperature ($T_{\rm K}$) of the molecular gas under the LTE assumption. The x-axis is the distance along the galactic plane from the galactic center.}\label{fig:ParaProfile}
\end{center}
\end{figure*}

$^{12}$CO~$J=1-0$, $J=2-1$, and $^{13}$CO~$J=1-0$ emission lines are clearly identified in most of the positions. Their integrated intensity ratios are quite informative of the physical conditions of the molecular gas in the galaxy. In this section, we present the $^{12}$CO/$^{13}$CO~$J=1-0$ and $^{12}$CO~$J=2-1$/$J=1-0$ line ratios (Fig.~\ref{fig:RatioProfile}) and some physical parameters derived from them (Fig.~\ref{fig:ParaProfile}; the parameters are summarized in Table~\ref{table:COlinepara}), after correcting the beam dilution at the corresponding line frequencies following the procedure described in \S\ref{sec:datareduction}.

The $^{12}$CO/$^{13}$CO line ratio is basically affected by two factors: the abundance of $^{13}$CO and the optical depth of the CO lines (e.g., \citealt{Alatalo15,JimenezDonaire17}). $^{13}$CO is preferentially formed in cold regions via ion-molecule interaction. This isotope-dependent fractionation will increase the $^{12}$CO/$^{13}$CO abundance ratio in warm molecular clouds. Furthermore, $^{13}$C is only produced by the CN cycle of Helium Burning in intermediate-mass stars, or as a secondary product in low-mass and high-mass ($>10\rm~M_\odot$) stars \citep{Sage91}. The total effect related to molecular abundance is thus a higher $^{12}$CO/$^{13}$CO abundance ratio in recent star forming regions (e.g., \citealt{JimenezDonaire17,Sliwa17a}).
On the other hand, the $^{12}$CO/$^{13}$CO line ratio could also be affected by the optical depth effects, which may either be caused by the change of gas temperature/density or by the presence of a diffuse, non-self-gravitating gas phase. These optical depth effects may result in the higher $^{12}$CO/$^{13}$CO line ratio in extensive star forming galaxies (e.g., \citealt{Tan11,JimenezDonaire17,Cormier18}). This is often explained as that $^{12}$CO emission largely arises from a warm diffuse inter-clumpy medium whereas $^{13}$CO emission mainly originates in denser cores. In active star forming regions, strong UV emission, SN shocks, and cosmic rays may heat, perturb, and dissociate the molecular cloud, resulting in a more significant decrease of the optical depth of the $^{12}$CO-rich envelope (than the $^{13}$CO-rich core). There are thus more $^{12}$CO photons escaping out of the molecular clouds, producing higher $^{12}$CO/$^{13}$CO line ratios. The $^{12}$CO/$^{13}$CO line ratios measured in NGC~5908 (with a median value of $\mathcal{R}_{\rm \mathsmaller{\mathsmaller{\frac{^{12}CO_{\rm 10}}{^{13}CO_{\rm 10}}}}}=7.0\pm1.7$; Table~\ref{table:COlinepara}) is in general comparable to some non-star-forming or normal star forming galaxies, but not as high as extreme starburst galaxies (e.g., \citealt{Tan11,Alatalo15,Sliwa17a,Sliwa17b,Cormier18}). Furthermore, $\mathcal{R}_{\rm \mathsmaller{\mathsmaller{\frac{^{12}CO_{\rm 10}}{^{13}CO_{\rm 10}}}}}$ does not increase in the nuclear region (Fig.~\ref{fig:RatioProfile}a). We therefore believe the $^{12}$CO/$^{13}$CO line ratio in NGC~5908 is not significantly affected by star formation feedback.

The $^{12}$CO~$J=2-1$/$J=1-0$ line ratio depends on the temperature and optical depth, thus the structure and heating sources of the molecular clouds. Based on the $^{12}$CO~$J=2-1$/$J=1-0$ line ratio, most of the Galactic molecular clouds could often be classified into two types: the low ratio gas with $\mathcal{R}_{\rm \mathsmaller{\mathsmaller{\frac{^{12}CO_{\rm 21}}{^{12}CO_{\rm 10}}}}}<0.7$ and the high ratio gas with $\mathcal{R}_{\rm \mathsmaller{\mathsmaller{\frac{^{12}CO_{\rm 21}}{^{12}CO_{\rm 10}}}}}=0.7-1.0$ \citep{Hasegawa97}. The low ratio gas is thought to be in molecular clumps with extended envelopes, where the CO emission mostly comes from low-density gas. On the other hand, the high ratio gas is in highly confined clumps with a steep density gradient and a thin CO emitting envelope, and is thus often found in the bright ridges of molecular clouds which are associated with star formation regions but not directly related to massive star formation (the low and high ratio gas is in general found in molecular clouds with a density $\lesssim10^2\rm~cm^{-3}$ or $\gtrsim10^2\rm~cm^{-3}$; e.g., \citealt{Penaloza17}). There is also another phase of molecular gas with $\mathcal{R}_{\rm \mathsmaller{\mathsmaller{\frac{^{12}CO_{\rm 21}}{^{12}CO_{\rm 10}}}}}>1$ (very high ratio gas). These gas clouds are influenced by significant external heating by UV photons from young stars or shock from SNe, both closely related to active star formation (e.g., \citealt{Hasegawa97}). 

The CO line ratios is in general highly affected by the local environmental conditions, such as the supersonic turbulence of the molecular clouds, the interstellar radiation field, and the cosmic ray ionization rate, etc. (e.g., \citealt{Penaloza18}). In most of the cases, efficient heating via these processes are only important in extreme starburst environments (e.g., \citealt{Papadopoulos12a}), although some molecular clouds with $\mathcal{R}_{\rm \mathsmaller{\mathsmaller{\frac{^{12}CO_{\rm 21}}{^{12}CO_{\rm 10}}}}}>1$ have also been suggested to be mainly heated by background cosmic rays (e.g., \citealt{Zhou18}). In most of the regions (except for the outermost regions with large error bars), the observed $^{12}$CO~$J=2-1$/$J=1-0$ line ratio in NGC~5908 is $<0.7$, with a median value of $\mathcal{R}_{\rm \mathsmaller{\mathsmaller{\frac{^{12}CO_{\rm 21}}{^{12}CO_{\rm 10}}}}}=0.43\pm0.11$ (Table~\ref{table:COlinepara}). Furthermore, we do not find any significant increase of $\mathcal{R}_{\rm \mathsmaller{\mathsmaller{\frac{^{12}CO_{\rm 21}}{^{12}CO_{\rm 10}}}}}$ in the nuclear region where the star formation is often the strongest in spiral galaxies. Therefore, we conclude that star formation in NGC~5908 is not intensive enough to modify $\mathcal{R}_{\rm \mathsmaller{\mathsmaller{\frac{^{12}CO_{\rm 21}}{^{12}CO_{\rm 10}}}}}$. However, we notice that there is a prominent $^{12}$CO~$J=2-1$ narrow line detected in the nuclear region (position ``0'' in Fig.~\ref{fig:12CO21spec}), but no corresponding component is detected in the $^{12}$CO~$J=1-0$ and $^{13}$CO~$J=1-0$ bands (Figs.~\ref{fig:12CO10spec}, \ref{fig:13CO10spec}, Table~\ref{table:COlinepara}), indicating the presence of some external heating sources. Either weak star formation or the AGN \citep{Li16b} may contribute in the heating of the molecular clouds in the nuclear region.

We next roughly estimate some derived physical parameters based on a few assumptions: (1) the molecular cloud is under the local thermal equilibrium (LTE) condition; (2) the emission lines from different CO isotopologues are co-spatial and have the same filling factor within the telescope beam, they also have the same excitation temperature; (3) the $^{13}$CO~$J=1-0$ line is optically thin while the $^{12}$CO~$J=1-0$ line is optically thick. All the assumptions need to be confirmed but are in general true for a regular galaxy with no extreme conditions. Under these assumptions, the optical depth of $^{13}$CO $\tau_{\rm ^{13}CO}$ can be derived from the $^{12}$CO/$^{13}$CO line ratio \citep{Tan11}:
\begin{equation}\label{equi:tau13CO}
\tau_{\rm ^{13}CO}=\ln[(1-\frac{I_{\rm ^{13}CO~J=1-0}}{I_{\rm ^{12}CO~J=1-0}})^{-1}].
\end{equation}
The derived $\tau_{\rm ^{13}CO}$ is listed in Table~\ref{table:COlinepara} and plotted in Fig.~\ref{fig:ParaProfile}a. $\tau_{\rm ^{13}CO}$ is $\lesssim0.2$ throughout the disk, with a median value of $0.16\pm0.03$.

The temperature of the molecular cloud could be estimated from the relative strength of the $^{12}$CO and $^{13}$CO lines. $N_{\rm H_2}$ can also be derived from the $^{13}$CO~$J=1-0$ line as \citep{Tan11}:
\begin{equation}\label{equi:NH213CO}
N_{\rm H_2}/{\rm cm^{-2}}=2.25\times10^{20}[\frac{\tau_{\rm ^{13}CO}}{1-e^{-\tau_{\rm ^{13}CO}}}]\frac{I_{\rm ^{13}CO~J=1-0}}{1-e^{-5.29/T_{\rm ex}}}
\end{equation}
where the $^{13}$CO abundance ${\rm [^{13}CO]/[H_2]}$ is taken to be $8\times10^{-5}/60$ \citep{Frerking82}. We also estimate the kinetic temperature of the gas $T_{\rm K}$ by equating Eq.~\ref{equi:NH213CO} to the $N_{\rm H_2}$ derived from $I_{{\rm ^{12}CO}~J=1-0}$. The derived $T_{\rm K}$ is listed in Table~\ref{table:COlinepara} and plotted in Fig.~\ref{fig:ParaProfile}b, with a median value of $32\pm9\rm~K$. At $r\lesssim8\rm~kpc$, $T_{\rm K}$ is quite stable at a value comparable to the median value, but $T_{\rm K}$ shows large uncertainty and fluctuation at larger radii. In interstellar molecular clouds without significant external heating sources, $T_{\rm K}$ is often $<20\rm~K$, so there could be moderate heating of the molecular clouds in NGC~5908, but the relatively low $T_{\rm K}$ compared to $\sim10^2\rm~K$ in starburst galaxies (e.g., \citealt{Tan11}) rules out a significant external heating by active star forming processes.

\subsection{Molecular gas rotation curve}\label{subsec:rotationcurve}

IRAM 30m observations of the molecular lines provide higher velocity resolution than any other gas components to study the fine structure of gas dynamics. The almost edge-on orientation of NGC~5908 is also optimized to study the rotation curve of this extremely massive spiral galaxy with least uncertainty on the inclination correction.

\begin{figure}
\begin{center}
\epsfig{figure=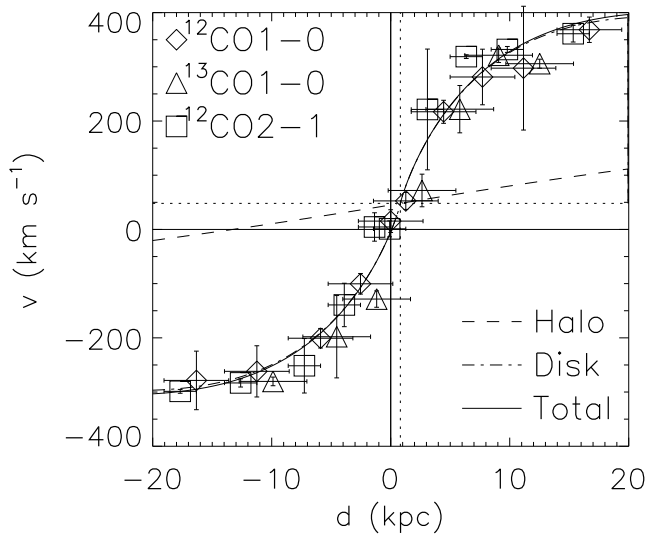,width=0.5\textwidth,angle=0, clip=}
\caption{Rotation curves of the $^{12}$CO~$J=1-0$ (diamond), $^{13}$CO~$J=1-0$ (triangle), and $^{12}$CO~$J=2-1$ (box) lines along the galactic disk. The x-axis is the distance along the galactic plane from the galactic center. Data points of the $^{13}$CO~$J=1-0$ and $^{12}$CO~$J=2-1$ lines are slightly shifted in the x-direction in order to avoid overlapping. The dotted vertical and horizontal lines show the best-fit shift from the optical center ($+0.8\rm~kpc$) and the systematic velocity of NGC~5908 ($+48\rm~km~s^{-1}$). The best-fit inclination angle is $89.98^\circ$, significantly larger than the value determined in optical ($65.31^\circ$, obtained from HyperLeda \citealt{Makarov14}). The solid curve is the universal rotation curve \citep{Salucci07} with a best-fit halo mass of $\log (M_{\rm vir}/M_\odot)=1.4\times10^{13}\rm~M_\odot$. The halo and disk components are plotted as the dashed and dash-dotted curves (very close to the solid curve), respectively.}\label{fig:rotationcurve}
\end{center}
\end{figure}

We combine the rotation curves of the $^{12}$CO~$J=1-0$, $^{13}$CO~$J=1-0$, and $^{12}$CO~$J=2-1$ lines in Fig.~\ref{fig:rotationcurve}. The average velocities of the lines and their errors, however, are measured based on averaging different components of the lines, as detailed in \S\ref{sec:datareduction}. At least at the outermost data points, we clearly see the trend of flattening of the rotation curves from the $^{12}$CO~$J=1-0$ and $J=2-1$ lines ($^{13}$CO~$J=1-0$ is not detected or covered), so the rotation curves could be used to estimate the dark matter halo mass of the galaxy (e.g., \citealt{Sofue01}).

We use the universal rotation curve of spiral galaxies from \citet{Salucci07} to characterize the shape of the rotation curves measured from the molecular lines. This universal rotation curve has two components: a stellar disk described by the Freeman disk \citep{Freeman70} and a dark matter halo described by the Burkert profile \citep{Burkert95} which converges to the NFW profile \citep{Navarro96} at large radii and often better fit the observed rotation curves. The universal rotation curve is thought to be valid out to the virial radius and has the virial mass of the dark matter halo ($M_{\rm vir}$) as the only galaxy indicator. With the inclination angle fixed at $i=65.31^\circ$ inferred from the optical morphology of the galaxy (obtained from HyperLeda; \citealt{Makarov14}), we cannot well fit the molecular line rotation curves ($\chi^2=383$ for a degree of freedom of $\rm d.o.f.=24$). We also notice that some authors have adopted a higher inclination angle of NGC~5908 (e.g., $i=81^\circ$ in \citealt{Kalinova17}). Therefore, we set it free and obtain a best-fit value of $i=(89.98_{-2.33}^{+2.36})^\circ$ (with a significantly reduced $\chi^2=145$ for $\rm d.o.f.=23$), indicating a perfectly edge-on molecular gas disk which may be slightly misaligned with the stellar disk. The best-fit virial mass of the dark matter halo is $M_{\rm vir}=1.42_{-0.08}^{+0.09}\times10^{13}\rm~M_{\rm \odot}$, roughly consistent with what is obtained in \citet{Li16b} ($M_{\rm 200}=8.1\times10^{12}\rm~M_\odot$) based on the \ion{H}{1} 21-cm line measurement (with a maximum rotation velocity of $347\rm~km~s^{-1}$, \citealt{vanMoorsel82}; consistent with our CO line measurement, Fig.~\ref{fig:rotationcurve}), considering the large systematical uncertainties in both model and measurement.

The above estimate of the dark matter halo mass based on the CO rotation curve may have large systematic uncertainties: (1) The peak of the universal rotation curve is always at $r\approx(10-20)\rm~kpc$ in a broad range of halo mass, within which the stellar component dominates \citep{Salucci07}. Without a good characterization at larger radii, the halo component cannot be directly constrained, especially for such massive galaxies whose dark matter distribution is very flat in the inner halo (the dashed curves in Fig.~\ref{fig:rotationcurve}). (2) For super-$L^\star$ galaxies as massive as NGC~5908, the stellar mass-halo mass relation may be quite different from those of lower mass galaxies (e.g., \citealt{Behroozi10}). Therefore, it may be problematic adopting $M_{\rm vir}$ as the only galaxy indicator. (3) \citet{Salucci07} does not include a bulge component in their rotation curve developed for spiral galaxies, while NGC~5908 has a significant bulge. (4) The back-reaction of baryons on the dark matter halo density distribution is not considered in the Burkert profile, and could make the radial distribution of dark matter significantly more concentrated (e.g., \citealt{Duffy10}), especially when the feedback is weak such as the case of NGC~5908 \citep{Li16b}. (5) NGC~5908 has a bright companion NGC~5905 located at $\approx200\rm~kpc$ away, whose tidal interaction may affect the dynamics of the gaseous disk. Considering these systematic uncertainties, we conclude that the dark matter halo mass of NGC~5908 is of the order of $\sim10^{13}\rm~M_{\rm \odot}$, one of the largest as a spiral galaxy. However, we emphasize that a more accurate measurement of $M_{\rm vir}$ requires better data and modeling. We will need a higher angular resolution and deeper CO line map to separate different overlaid velocity components and to detect gas at larger radii. We also need an updated analytical rotation curve including a bulge component and adopting the stellar mass-halo mass relation for super-$L^\star$ galaxies.

\subsection{How does cold gas affect the gas budget, star formation efficiency, and feedback efficiency?}\label{subsec:gasbudget}

In this section, we discuss how the measured cold gas properties help us to understand the evolution of this extremely massive, isolated spiral galaxy. We will compare the measured molecular gas content (plus the atomic gas mass obtained from archive) to the SFR, hot CGM radiative cooling rate ($\dot{M}_{\rm cool}$), and the expected baryon mass obtained from multi-wavelength observations. As the molecular gas mass estimated from both the ${\rm ^{12}CO}~J=1-0$ and ${\rm ^{13}CO}~J=1-0$ lines have significant systematic uncertainties, we adopt an average value of them ($M_{\rm H_2}=6.8\pm1.7\times10^{9}\rm~M_\odot$) in the following discussions. We also adopt an atomic hydrogen gas mass of $M_{\rm HI}=5.9\times10^{9}\rm~M_\odot$ from the \ion{H}{1} 21-cm line observations of \citet{vanMoorsel82}, so the total cold gas mass of NGC~5908 is $M_{\rm H_2+HI}\approx1.3\times10^{10}\rm~M_\odot$.

Galaxies in the local universe are observed to have the baryon content much lower than is needed in Big Bang nucleosynthesis, forming the ``missing baryon'' problem (e.g., \citealt{Bregman07,Bregman18,Li18}). The total mass of cold molecular and atomic gas of NGC~5908 is in general comparable to the mass contained in the extended hot CGM ($M_{\rm hot}=1.4_{-0.6}^{+3.3}\times10^{10}\rm~M_\odot$, Table~\ref{table:NGC5908}), so is important in the gas budget. However, compared to the stellar mass of $M_*=2.6\times10^{11}\rm~M_\odot$ and the expected total baryon mass of $\approx2\times10^{12}\rm~M_\odot$ from the estimated dark matter halo mass \citep{Li17}, the mass contained in the multi-phase gases is not quite important and far from accounting for the missing baryons.

The relatively small gas content, however, may still be important in the gas circulation and star formation of this massive, isolated, spiral galaxy.
We first check if the galaxy follows the Kennicutt-Schmidt law or not [$\Sigma_{\rm SFR}/{\rm M_\odot~yr^{-1}~kpc^{-2}}=(2.5\pm0.7)\times10^{-4}(\Sigma_{\rm H_2+HI}/{\rm M_\odot~pc^{-2}})^{1.4\pm0.15}$; \citealt{Kennicutt98}]. Assuming the radius of the star forming disk to be $r=20\rm~kpc$, consistent with the size of the dustlane and the detected CO lines (also see CARMA observations of the ${\rm ^{12}CO}~J=1-0$ line in \citealt{Bolatto17}), we can obtain the molecular gas surface density to be $\Sigma_{\rm H_2}=5.4\pm1.4\rm~M_\odot~pc^{-2}$, or the total cold gas surface density to be $\Sigma_{\rm H_2+HI}=10.1\pm1.4\rm~M_\odot~pc^{-2}$. Adopting the Kennicutt-Schmidt law, the expected SFR surface density would be $\Sigma_{\rm SFR,expect}=6.0_{-2.3}^{+3.3}\times10^{-3}\rm~M_\odot~yr^{-1}~kpc^{-2}$ (or $2.5_{-1.2}^{+1.7}\times10^{-3}\rm~M_\odot~yr^{-1}~kpc^{-2}$ if only considering the molecular gas traced by CO lines). Considering the large systematic uncertainties in the measurement of cold gas mass and the scatter of the Kennicutt-Schmidt law, this value is in general comparable to the measured $\Sigma_{\rm SFR}$ of $\approx3\times10^{-3}\rm~M_\odot~yr^{-1}~kpc^{-2}$ (\S\ref{sec:Introduction}; \citealt{Li16b}). We therefore do not find a significant suppression of star formation in NGC~5908, as often suggested in disk galaxies with a huge bulge due to some star formation quenching mechanisms such as morphological quenching (e.g., \citealt{Martig09,Li09,Li11}).

Accretion of the radiatively cooled hot halo gas may be an important source of cold gas observed in massive galaxies with an extensive hot CGM \citep{Li17,Li18}. Based on deep \emph{XMM-Newton} X-ray observations, we have measured the radiative cooling rate of a sample of massive spiral galaxies with the virial temperature high enough to produce virialized soft X-ray emitting gas, including NGC~5908 \citep{Li16b,Li17}. We obtained a cooling rate of the hot CGM within the cooling radius of $\dot{M}_{\rm cool}=0.37~(<1.55)\rm~M_\odot~yr^{-1}$ for NGC~5908, or the total cooled gas mass of $M_{\rm cool}<1.5\times10^{10}\rm~M_\odot$ within the Hubble time. The large uncertainty is caused by a fairly X-ray bright AGN (difficult to be cleanly removed from the \emph{XMM-Newton} image) which strongly affect the characterization of the radial distribution of hot gas in the inner halo. $\dot{M}_{\rm cool}$ is about one order of magnitude lower than the SFR, so in principle, even if all the cooled gas can be used for star formation, the accretion of the cooled hot CGM is still insufficient to compensate the gas consumed in star formation. It is not impossible to have other gas supplies such as the accretion of cool gas from the companion galaxy NGC~5905. However, the HI 21-cm line observation of NGC~5905/5908 did not show any signatures of gas transfer \citep{vanMoorsel82}. We therefore conclude that the cold gas observed in NGC~5908 is mostly leftover gas from previous starburst, instead of newly acquired from the very isolated environment.

Assuming the current SFR, it will need about ten times of the Hubble time to accumulate the huge stellar mass of NGC~5908. Therefore, the galaxy must be in a stage(s) before with much higher growth rate, either from mergers or starbursts. On the other hand, the SFR of NGC~5908 is relatively high among the CGM-MASS galaxies \citep{Li16b}, and is about one order of magnitude higher than that of M31 ($\sim0.4\rm~M_\odot~yr^{-1}$; \citealt{Barmby06}), another massive quiescent spiral galaxy in our close neighbourhood. Since there is no signature of recent triggering of star formation (such as galaxy merger, gas transfer, or enhanced cooling of the CGM etc.), we conclude that NGC~5908 is not completely quiescent, and it may be in the early stage after the previous starburst, before the star formation is completely extinguished and the ejected gas cools and falls back to the galaxy.

SN feedback is a key ingredient regulating the galaxy evolution. The thermalization efficiency, or the fraction of SN energy released as multi-wavelength emissions in the ISM or the CGM, is often only a few percent (e.g., \citealt{Li13b,Walch15}). Many factors may affect the thermalization efficiency (e.g., \citealt{Wang16}), which is poorly constrained directly from observations (e.g., \citealt{Strickland09}). A key factor which may affect the observed X-ray radiation efficiency (a good tracer of feedback or thermalization efficiency) is the cold-hot gas interaction, which is often expected to enhance the observed X-ray emission via many micro-processes at the interface of different gas phases (such as turbulent mixing, thermal evaporation, charge exchange, etc.; e.g., \citealt{Strickland02,Li13b,Zhang14}). NGC~5908 and most other CGM-MASS galaxies are among the spiral galaxies with the lowest X-ray radiation efficiency \citep{Li17}. However, unlike other CGM-MASS galaxies which are mostly poor in cool gas (Li J., et al., in prep.) so the low X-ray radiation efficiency is easily explained as the result of the lack of gas mixing (e.g., \citealt{Tang09}), NGC~5908 is quite rich in cool gas and is often expected to be relatively X-ray bright (e.g., \citealt{Li09,Li15}). The low X-ray radiation efficiency provides evidence that the previously internally ejected or externally accreted gas in the CGM is not yet cooled and well mixed with the cold gas.

\section{Summary and Conclusion}\label{sec:summary}

We present IRAM 30m observations of the $^{12}$CO~$J=1-0$, $^{12}$CO~$J=2-1$, and $^{13}$CO~$J=1-0$ lines from the highly inclined massive spiral galaxy NGC~5908 selected from the CGM-MASS sample. All these three lines have been detected in most of the positions along the galactic disk, except that $^{13}$CO~$J=1-0$ is not covered or clearly detected at the outermost positions. We model the lines with one or two gaussian components and calculate the integrated line intensities, line ratios, average velocity, and a few derived physical parameters. Below we summarize our major results and conclusions.

$\bullet$ The total mass of molecular gas in NGC~5908 estimated from the integrated $^{12}$CO~$J=1-0$ ($^{13}$CO~$J=1-0$) line intensity is $M_{\rm H_2}=8.3\pm0.4\times10^{9}\rm~M_\odot$ ($5.2\pm0.4\times10^{9}\rm~M_\odot$). Adding the atomic component measured with the \ion{H}{1} 21-cm line from the literature, the total cool gas mass is $\approx1.3\times10^{10}\rm~M_\odot$. The cool gas mass is thus comparable to the total mass of hot gas in the halo, so makes a significant contribution to the galaxy's gas budget, but is still far from sufficient to account for the ``missing baryons''.

$\bullet$ The rotation curves determined with a combination of different molecular lines significantly flatten at $r\approx15\rm~kpc$. Fitting the measured velocities of different molecular lines with the universal rotation curve from \citet{Salucci07} indicates that NGC~5908 has a dark matter halo with the mass of $M_{\rm vir}\sim10^{13}\rm~M_{\rm \odot}$. Although better data and modeling are required to accurately measure the halo mass, NGC~5908 is definitely among the most massive spiral galaxies yet known in the local universe. The basic scenario of gas cycling and star formation properties may thus be representative of a massive spiral galaxy evolved largely in isolation.

$\bullet$ The $^{12}$CO/$^{13}$CO~$J=1-0$ line ratio is in general comparable to normal star forming galaxies, but not as high as extremely starburst ones. It is thus likely that star formation feedback does not seriously affect the molecular gas properties. In the mean time, the $^{12}$CO~$J=2-1$/$J=1-0$ line ratio and the estimated gas temperature also indicate relatively weak star formation in this massive spiral galaxy. Comparison to the star formation law indicates that NGC~5908 is converting cool gas to young stars at a normal efficiency, with no significant suppression of star formation as often suggested in disk galaxies with a huge bulge. The radiative cooling of the hot CGM or any other external gas supply are probably insufficient to compensate the gas consumed in star formation. The galaxy must have a fast growth stage with mergers and/or starbursts in the past. It is probably now at an early stage after the starburst, with plenty of leftover cool gas, a relatively high SFR, low hot CGM cooling rate, and low X-ray emissivity.

\bigskip
\noindent\textbf{\uppercase{acknowledgements}}
\smallskip\\
\noindent The authors acknowledge Dr. Zhi-Yu Zhang and Dr. Qinghua Tan for helpful discussions. JTL and JNB acknowledge the financial support from NASA through the grants NNX15AM93G, SOF05-0020, 80NSSC18K0536, and NAS8-03060. PZ acknowledges the support from the NWO Veni Fellowship, grant no. 639.041.647 and NSFC grants 11503008 and 11590781. XJJ and YG are partially supported by National Key Basic Research and Development Program of China (Grant No. 2017YFA0402704), NSFC Grant No. 11420101002, and Chinese Academy of Sciences Key Research Program of Frontier Sciences (Grant No. QYZDJ-SSW-SLH008).

\clearpage

\begin{table*}
\begin{center}
\rotatebox{90}{
\begin{minipage}{\textheight}
\caption{Observed and Derived Properties of the CO Lines}
\footnotesize
\tabcolsep=3.0pt%
\begin{tabular}{lcccccccccccccccccccc}
\hline
Region & $d$ & $I_{\rm ^{12}CO_{\rm 10}}$ & $v_{\rm ^{12}CO_{\rm 10}}$ & $I_{\rm ^{13}CO_{\rm 10}}$ & $v_{\rm ^{13}CO_{\rm 10}}$ & $I_{\rm ^{12}CO_{\rm 21}}$ & $v_{\rm ^{12}CO_{\rm 21}}$ & $\mathcal{R}_{\rm \mathsmaller{\mathsmaller{\frac{^{12}CO_{\rm 10}}{^{13}CO_{\rm 10}}}}}$ & $\mathcal{R}_{\rm \mathsmaller{\mathsmaller{\frac{^{12}CO_{\rm 21}}{^{12}CO_{\rm 10}}}}}$ & $N_{\rm H_2,^{12}CO}$ & $N_{\rm H_2,^{13}CO}$ & $\tau(\rm ^{13}CO)$ & $T_{\rm K}$ \\
 & $^{\prime\prime}$ & $\rm K~km~s^{-1}$ & $\rm km~s^{-1}$ & $\rm K~km~s^{-1}$ & $\rm km~s^{-1}$ & $\rm K~km~s^{-1}$ & $\rm km~s^{-1}$ & & & $10^{21}\rm~cm^{-2}$ & $10^{21}\rm~cm^{-2}$ &  & K \\
\hline
0 (avg) &   5.04 & $28.03\pm 3.47$ & $52.6\pm 15.7$ & $3.26\pm 0.51$ & $71.9\pm 30.1$ & $30.68\pm 2.42$ & $1.1\pm 14.8$ & $7.53\pm 1.51$ & $0.44\pm 0.06$ & $5.61\pm 0.69$ & $3.26\pm 0.51$ & $0.142\pm 0.031$ & $34.9\pm 7.6$ \\
0 (nrw) & - & $5.80\pm 2.48$ & $79.9\pm 9.2$ & - & - & $4.19\pm 1.05$ & $10.3\pm 4.3$ & - & - & - & - & - & - \\
0 (brd) & - & $22.23\pm 2.44$ & $45.5\pm 14.6$ & - & - & $26.49\pm 2.18$ & $-0.3\pm 17.1$ & - & - & - & - & - & - \\
1 (avg) &  17.64 & $23.01\pm 1.45$ & $217.0\pm 21.2$ & $2.92\pm 0.42$ & $221.8\pm 43.7$ & $21.99\pm 7.38$ & $221.4\pm 111.4$ & $6.98\pm 1.09$ & $0.42\pm 0.14$ & $4.60\pm 0.29$ & $2.92\pm 0.42$ & $0.155\pm 0.026$ & $31.6\pm 5.4$ \\
1 (nrw) & - & $5.52\pm 0.88$ & $315.0\pm 3.1$ & $0.81\pm 0.30$ & $87.2\pm 23.1$ & $8.60\pm 5.03$ & $297.3\pm 12.9$ & - & - & - & - & - & - \\
1 (brd) & - & $17.49\pm 1.15$ & $186.1\pm 7.4$ & $2.11\pm 0.29$ & $273.8\pm 8.7$ & $13.38\pm 5.40$ & $172.6\pm 33.6$ & - & - & - & - & - & - \\
2 (avg) & -10.08 & $25.05\pm 2.19$ & $-100.3\pm 19.0$ & $3.36\pm 0.43$ & $-128.5\pm 15.1$ & $22.61\pm 3.86$ & $-139.3\pm 40.1$ & $6.60\pm 1.02$ & $0.39\pm 0.08$ & $5.01\pm 0.44$ & $3.36\pm 0.43$ & $0.164\pm 0.028$ & $29.6\pm 5.0$ \\
2 (nrw) & - & $3.65\pm 1.37$ & $-254.4\pm 9.2$ & - & - & $4.99\pm 2.47$ & $-251.2\pm 10.8$ & - & - & - & - & - & - \\
2 (brd) & - & $21.39\pm 1.71$ & $-74.0\pm 9.3$ & - & - & $17.62\pm 2.97$ & $-107.6\pm 11.6$ & - & - & - & - & - & - \\
3 (avg) &  30.60 & $12.85\pm 1.63$ & $281.3\pm 51.3$ & $1.08\pm 0.28$ & $321.3\pm 13.4$ & $9.28\pm 0.88$ & $318.8\pm 3.4$ & $10.50\pm 3.01$ & $0.43\pm 0.07$ & $2.57\pm 0.33$ & $1.08\pm 0.28$ & $0.100\pm 0.030$ & $50.4\pm 15.2$ \\
3 (nrw) & - & $5.39\pm 1.05$ & $330.2\pm 2.2$ & - & - & - & -& - & - & - & - & - & - \\
3 (brd) & - & $7.46\pm 1.25$ & $246.0\pm 11.4$ & - & - & - & -& - & - & - & - & - & - \\
4 (avg) & -23.40 & $22.35\pm 1.33$ & $-200.7\pm 18.0$ & $3.02\pm 0.78$ & $-197.9\pm 76.0$ & $13.25\pm 1.79$ & $-251.5\pm 50.3$ & $6.45\pm 1.72$ & $0.30\pm 0.04$ & $4.47\pm 0.27$ & $3.02\pm 0.78$ & $0.168\pm 0.049$ & $29.3\pm 8.6$ \\
4 (nrw) & - & $6.36\pm 0.80$ & $-281.1\pm 2.5$ & $1.04\pm 0.45$ & $-279.6\pm 9.2$ & $1.71\pm 1.06$ & $-308.2\pm 4.4$ & - & - & - & - & - & - \\
4 (brd) & - & $15.98\pm 1.06$ & $-168.6\pm 5.4$ & $1.97\pm 0.64$ & $-154.8\pm 25.9$ & $11.53\pm 1.44$ & $-243.1\pm 9.3$ & - & - & - & - & - & - \\
5 (avg) &  44.28 & $3.71\pm 0.99$ & $297.6\pm 114.2$ & $0.55\pm 0.17$ & $305.8\pm 8.3$ & $3.89\pm 0.70$ & $332.1\pm 5.5$ & $6.25\pm 2.51$ & $0.68\pm 0.22$ & $0.74\pm 0.20$ & $0.55\pm 0.17$ & $0.174\pm 0.076$ & $26.5\pm 11.8$ \\
5 (nrw) & - & $1.13\pm 0.68$ & $335.0\pm 3.2$ & - & - & - & -& - & - & - & - & - & - \\
5 (brd) & - & $2.57\pm 0.72$ & $281.1\pm 11.8$ & - & - & - & -& - & - & - & - & - & - \\
6 (avg) & -44.64 & $7.51\pm 0.93$ & $-261.8\pm 47.3$ & $0.80\pm 0.16$ & $-280.4\pm 8.0$ & $4.79\pm 0.77$ & $-283.1\pm 7.6$ & $8.45\pm 2.01$ & $0.34\pm 0.07$ & $1.50\pm 0.19$ & $0.80\pm 0.16$ & $0.126\pm 0.032$ & $38.7\pm 9.9$ \\
6 (nrw) & - & $1.84\pm 0.60$ & $-301.6\pm 2.8$ & - & - & - & -& - & - & - & - & - & - \\
6 (brd) & - & $5.67\pm 0.72$ & $-248.9\pm 8.9$ & - & - & - & -& - & - & - & - & - & - \\
7 (avg) &  66.24 & $1.33\pm 0.37$ & $368.4\pm 23.4$ & - & - & $2.43\pm 0.60$ & $361.1\pm 15.4$ & - & $0.71\pm 0.27$ & $0.27\pm 0.07$ & - & - & - \\
8 (avg) &   0.00 & $28.38\pm 6.75$ & $15.4\pm 21.0$ & - & - & $27.63\pm 2.93$ & $4.5\pm 26.2$ & - & $0.41\pm 0.11$ & $5.68\pm 1.35$ & - & - & - \\
8 (nrw) & - & $3.25\pm 3.94$ & $11.9\pm 28.8$ & - & - & $21.04\pm 1.25$ & $17.2\pm 4.1$ & - & - & - & - & - & - \\
8 (brd) & - & $25.13\pm 5.47$ & $15.8\pm 22.7$ & - & - & $6.59\pm 2.65$ & $-35.9\pm 108.1$ & - & - & - & - & - & - \\
9 (avg) & -64.80 & $1.09\pm 0.15$ & $-278.6\pm 54.0$ & - & - & $0.55\pm 0.24$ & $-298.7\pm 3.5$ & - & $0.49\pm 0.22$ & $0.22\pm 0.03$ & - & - & - \\
9 (nrw) & - & $0.24\pm 0.09$ & $-236.2\pm 3.1$ & - & - & - & -& - & - & - & - & - & - \\
9 (brd) & - & $0.85\pm 0.12$ & $-290.6\pm 2.8$ & - & - & - & -& - & - & - & - & - & - \\
\hline
\end{tabular}\label{table:COlinepara}\\
$d$ is the distance along the galactic plane from the center of the galaxy (northwest as positive). Some lines from some regions are fitted with a 2-gauss model, so there is a narrow component (nrw) and a broad component (brd). The average (avg) line velocity $v$ of the two components are weighted by their integrated line intensity $I$. For the lines fitted with just one gaussian component, the average $v$ is the directly measured value. The integrated line intensity $I$ has been corrected for the main beam and forward efficiencies in the $J=1-0$ and $J=2-1$ bands, respectively. The line ratio $\mathcal{R}$ is further corrected for beam dilution. $N_{\rm H_2,^{12}CO}$ and $N_{\rm H_2,^{13}CO}$ are the molecular gas column density derived from $I_{\rm ^{12}CO~J=1-0}$ and $I_{\rm ^{13}CO~J=1-0}$, respectively. $\tau(\rm ^{13}CO)$ is the optical depth at $^{13}$CO derived from Equ.~\ref{equi:tau13CO}. $T_{\rm K}$ is the kinetic temperature derived from $N_{\rm H_2}$ and Eq.~\ref{equi:NH213CO}. All errors are quoted as 1~$\rm\sigma$ confidence level.
\end{minipage}
}
\end{center}
\end{table*}

\end{document}